\documentclass[useAMS,usenatbib]{mn2e}

\usepackage{natbib}
\usepackage{graphicx}
\usepackage{graphics}
\usepackage[T1]{fontenc}
\usepackage{times}

\newcommand{\chem}{\everymath={\fam0 }\fam0 }  
\newcommand{\iso}[2]{{}^{#1}_{#2}}

\title[Entrainment and jet-related emission in Centaurus\,A]{Internal entrainment and the origin of jet-related broad-band emission in Centaurus\,A}
\author[Sarka Wykes et al.]{Sarka Wykes$^{1,2}$\thanks{E-mail: sarka@astro.ru.nl}, Martin J. Hardcastle$^{3}$\thanks{E-mail: m.j.hardcastle@herts.ac.uk}, Amanda I. Karakas$^{4}$ and Jorick S. Vink$^{5}$
\\
$^{1}$Department of Astrophysics/IMAPP, Radboud University Nijmegen, PO Box 9010, 6500 GL Nijmegen, The Netherlands\\
$^{2}$Anton Pannekoek Institute for Astronomy, University of Amsterdam, PO Box 94249, 1090 GE Amsterdam, The Netherlands\\
$^{3}$School of Physics, Astronomy and Mathematics, University of Hertfordshire, College Lane, Hatfield, Hertfordshire AL10 9AB, UK\\
$^{4}$Research School of Astronomy and Astrophysics, Australian National University, Canberra, ACT 2611, Australia\\
$^{5}$Armagh Observatory, College Hill, Armagh, BT61 9DG, Northern Ireland, UK
}

\begin{document}

\date{Accepted 2014 November 14. Received 2014 November 10; in original form 2014 September 18}

\pagerange{\pageref{firstpage}--\pageref{lastpage}} \pubyear{2014}

\maketitle

\label{firstpage}

\begin{abstract}
The dimensions of Fanaroff-Riley class I jets and the stellar densities at galactic centres imply that there will be numerous interactions between the jet and stellar winds. These may give rise to the observed diffuse and `knotty' structure of the jets in the X-ray, and can also mass load the jets. We performed modelling of internal entrainment from stars intercepted by Centaurus\,A's jet, using stellar evolution- and wind codes. From photometry and a code-synthesised population of $12$\,Gyr ($Z=0.004$), $3$\,Gyr ($Z=0.008$) and $0-60$\,Myr ($Z=0.02$) stars, appropriate for the parent elliptical NGC\,5128, the total number of stars in the jet is $\sim8\times10^8$. Our model is energetically capable of producing the observed X-ray emission, even without young stars. We also reproduce the radio through X-ray spectrum of the jet, albeit in a downstream region with distinctly fewer young stars, and recover the mean X-ray spectral index. We derive an internal entrainment rate of $\sim2.3\times10^{-3}$\,M$_{\odot}$\,yr$^{-1}$ which implies substantial jet deceleration. Our absolute nucleosynthetic yields for the AGB stellar population in the jet show the highest amounts for $\chem \iso{4}{}He$, $\chem \iso{16}{}O$, $\chem \iso{12}{}C$, $\chem \iso{14}{}N$ and $\chem \iso{20}{}Ne$. If some of the events at $\ge55$\,EeV detected by the Pierre Auger Observatory originate from internal entrainment in Centaurus\,A, we predict that their composition will be largely intermediate-mass nuclei with $\chem \iso{16}{}O$, $\chem \iso{12}{}C$ and $\chem \iso{14}{}N$ the key isotopes.

\end{abstract}


\section{Introduction} \label{sect:introduction}

A growing body of opinion suggests that jets in Fanaroff-Riley class I (FR\,I; \citealp{FAN74}) radio galaxies start off essentialy leptonic and become mass loaded through entrainment, causing them to decelerate on kiloparsec scales \citep{REY96, LAI02, HUB06, WYK13, PER14a}. Mass loading of baryons comes about through various routes. While a number of authors (\citealp{YOU86}; \citealp{BIC84, BIC94}; \citealp{LAI02, WYK13}) have focused mostly on external entrainment from hot gas of the surrounding galaxy ISM, more recent literature (\citealp{KOM94, BOW96, BED97, LAI02, HUB06, BOS12, WYK13, HUA13, PER14a, PER14b}) considers internal entrainment, i.e., mass loading via stellar winds from stars contained within the jet. The contribution from internal entrainment, in FR\,I sources where both processes are likely to operate, appears to be in excess of that from external entrainment in the inner regions of the galaxy \citep{LAI02, WYK13, PER14b}.\footnote{The \cite{LAI02} results on 3C\,31 in particular are consistent with all of the mass input within $1$\,kpc from the nucleus being due to stellar mass loss; however, they cannot exclude the possibility of external entrainment within this distance.}

Stars in elliptical galaxies are predominantly from old populations, with K- and M-type stars dominating the galaxy's integrated light (e.g. \citealp{ATH02, VDO10}) and previous attempts to model stellar mass input into jets assume mostly K and M stars, as above \citep{KOM94, BOW96}. A young stellar component is expected in some sources through the presence of a starburst. Central stellar densities for massive ellipticals can be of the order $10^{10}$\,M$_{\odot}$\,kpc$^{-3}$ \citep{TRE04} although they fall off rapidly with distance from the centre. Hence, inevitably, there are stars in the path of the jet. However, the jet interacts not with the star itself but with the stellar wind. Stellar winds, with a plethora of wind-driving mechanisms including radiative-line driving, dust driving, Alfv\'en wave pressure, stellar rotation and stellar pulsation (e.g. \citealp{LAM99, NEI13}), occur all across the main sequence and beyond until the onset of the compact object phase. The stellar wind typically extends for tens to hundreds of stellar radii. The stages of dramatically enhanced mass-loss rates during the evolution include, for low-mass stars, Asymptotic Giant Branch (AGB) phase, i.e. voluminous stars with slow, dense winds and a mass-loss rate of $\dot{M}\sim1\times10^{-5}-1\times10^{-4}$\,M$_{\odot}$\,yr$^{-1}$ (e.g. \citealp{VAS93, VLO99, OLO02}). Among high-mass stars, the Luminous Blue Variable (LBV) phase with slow, very dense winds lose typically $\dot{M}\sim1\times10^{-4}-1\times10^{-2}$\,M$_{\odot}$\,yr$^{-1}$ (e.g. \citealp{VIN02}), and the follow-up phase,\footnote{But see \cite{KOT06} and \cite{SMI14b} who disfavour a scenario with LBVs being massive stars in transition to WR stars.} the Wolf-Rayet (WR) stars -- a stage with high-velocity winds and $\dot{M}\sim1\times10^{-5}-1\times10^{-4}$\,M$_{\odot}$\,yr$^{-1}$ (e.g. \citealp{SAN12}). Considerable uncertainties exist in mass-loss rates for high-mass stars, particularly due to wind clumping on small spatial scales (e.g. \citealp{FUL06, VIN12}). The wind strength is generally expressed as the ratio of terminal velocity to escape velocity $v_{\infty}/v_{\rm esc}$, which is a strong function of the effective temperature $T_{\rm eff}$. The properties of the high-mass-loss stars are summarized in Table\,\ref{tab:para}.

Centaurus\,A is the closest ($3.8\pm0.1$\,Mpc; \citealp{HARR10a}) radio galaxy, a FR\,I object with a physical age of $\sim560$\,Myr \citep{WYK13, EIL14, WYK14}, associated with the massive Ep galaxy NGC\,5128 \citep{HARR10b}. The nucleus harbours a supermassive black hole, with a mass $M_{\rm BH} = (5.5\pm3.0)\times10^7$\,M$_{\odot}$ derived by \cite{CAP09} from stellar kinematics. A prominent dust lane (e.g. \citealp{DUF79, EBN83, ECK90}), with starburst (e.g. \citealp{MOL81, UNG00, MIN04}), crosses the central parts. Due to its luminosity and proximity, Centaurus\,A furnishes a means of testing models of jet energetics, particle content, particle acceleration, and the evolution of low-power radio galaxies in general. 

\begin{figure}
\includegraphics[width=\linewidth]{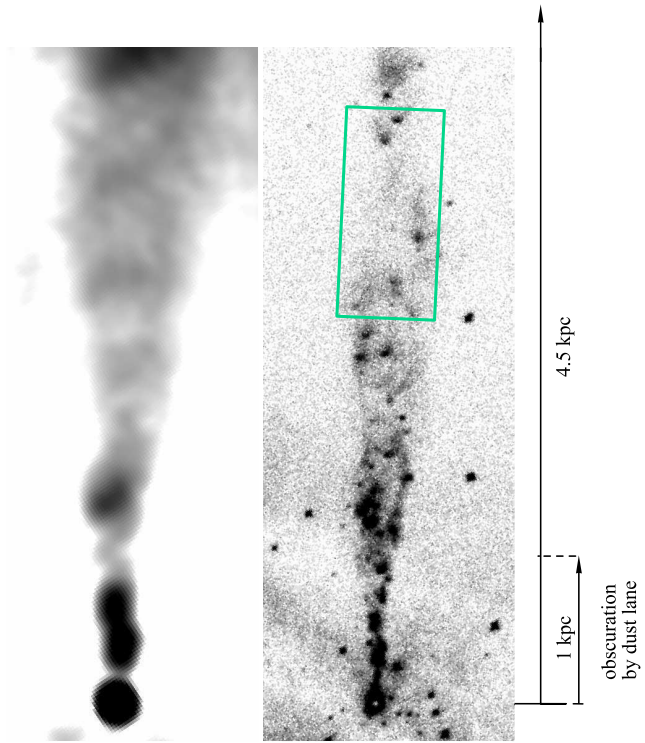}
\caption{Left: Very Large Array (VLA) 5\,GHz radio map at $6$\,arcsec resolution showing the emission of the $\sim4.5$\,kpc-scale (projected) jet. Right: {\it Chandra} X-ray image at $1$\,arcsec resolution of the same part of the jet. Clearly distinguishable X-ray knots are visible throughout the jet. Also shown is the extent of the dust lane, and the `inner' region (green box) considered by \citet{HAR06} and discussed in Sections\,\ref{sect:spectrum_intro} and \ref{sect:spectrum_results}. Figure adapted from \citet{HAR11}.} \label{fig:fig1}
\end{figure}

Centaurus\,A shows a twin jet in radio and X-ray bands, symmetrical on parsec scales but with evident asymmetry on kpc scales. The main (i.e. northern) jet (Fig.\,\ref{fig:fig1}) which is markedly brighter than the counterjet, is seen at a viewing angle of approximately 50$^\circ$ \citep{TIN98, HAR03}. From photoionization models for such a viewing angle, \cite{BIC13} have derived the Lorentz factor of the jet $\Gamma_{\rm j}\la5$. \cite{TIN01} and \cite{MUL14} have measured jet component speeds of $0.1-0.3c$ at subparsec scales, while \cite{HAR03} found projected speeds of $0.5c$ at $\sim100$\,pc, pointing towards jet acceleration downstream or to sampling of disparate jet layers. The power of the jet driving the currently active $\sim5$\,kpc-scale lobes (i.e. the inner lobes) is well constrained, $P_{\rm j}\sim1\times10^{43}$\,erg\,s$^{-1}$ (\citealp{WYK13} and references therein); the physical age of this jet is probably $\sim2$\,Myr \citep{CRO09, WYK13}. A large number of radio and X-ray knots is discernible in the jet on kpc scales \citep{KRA02, HAR03, KAT06, WOR08, TIN09, GOO10}, with the radio knots of larger proper motions showing comparatively little X-ray emission \citep{GOO10}. In addition to the knots, diffuse X-ray emission extends out to about $4.5$\,kpc in projection from the nucleus \citep{HAR07}. The spectral indices of individual X-ray knots have a wide range of values, but at least some have flat spectra suggestive of ongoing particle acceleration \citep{GOO10}; in principle, ongoing particle acceleration is required for any part of the jet that produces X-ray emission. The overall spectrum of diffuse emission in the jet is flat in the radio but steepens before the mid-infrared, with the integrated X-ray spectral index being steeper than expected from simple continuous-injection models \citep{HAR06}. The spectral index of diffuse emission is also a function of distance along the jet, with steeper spectra being observed at larger distances \citep{HAR07}.

\cite{HAR03} and \cite{NUL10} suggested that some or all of the knots might be the result of the interaction between the jet and the winds of high-mass-loss stars. An association with supernova remnants has been disfavoured \citep{HAR03} based on such supernova remnants' expected emission mechanism (thermal) and relatively low temperature ($\la1$\,keV).
\begin{table*}
\begin{center}
\caption{Properties of high-mass-loss stars: initial stellar mass $M_{\rm init}$, stage duration $t_{\rm stage}$, mass-loss rate at that stage $\dot{M}$, total particle number density of the wind $n_{\rm w}$ (from the continuity equation), stellar mass $M_*$, stellar radius $R_*$ (in optical and X rays), effective temperature $T_{\rm eff}$, X-ray luminosity $L_{\rm X}$, and the ratio of terminal velocity to escape velocity $v_{\infty}/v_{\rm esc}$.}
\label{tab:para}
\begin{tabular}{lccccccccc}
\hline \\ [-1.5ex]
   & $M_{\rm init}$ & $t_{\rm stage}$ & $\dot{M}$ & $n_{\rm w}$ & $M_*$ & $R_*$ & $T_{\rm eff}$ & $L_{\rm X}$ & $v_{\infty}/v_{\rm esc}$\\
   & (M$_{\odot}$) & (kyr) & (M$_{\odot}$\,yr$^{-1}$) & (cm$^{-3}$) & (M$_{\odot}$) & (R$_{\odot}$) & (K) & (erg\,s$^{-1}$) &\\
[-1.5ex] 
\multicolumn{7}{l}{} \\
\hline \\ [-1.5ex]
AGB  & $0.8 - 8.0$ & $100-1000$ & $1\times10^{-5} - 1\times10^{-4}$ & $10-10^{11}$ & $0.5 - 5$ & $100 - 1000$ & $2000 - 4000$ &  & $0.1$ \\
LBV  & $\ge30$     & $10 - 100$ & $1\times10^{-4} - 1\times10^{-2}$ & $1-10^{12}$ & $10 - 50$ & $20 - 100$ & $10\,000 - 30\,000$ & $1\times10^{31}$ & $0.2$ \\
WR   & $\ge20$     & $10 - 100$ & $1\times10^{-5} - 1\times10^{-4}$ & $10^{-4}-10^{14}$ & $5 - 10$ & $2 - 3$ & $50\,000 - 150\,000$ & $1\times10^{31}$ & $3$ \\
[1ex] \hline
\end{tabular} \\
\begin{tabular}{l}
\scriptsize{Note. The lower limit on $n_{\rm w}$ represents the number density at the stand-off distance, the upper limit the number density at the stellar surface. The escape velocity includes the Eddington factor $\Gamma_{\rm E}$.}
\end{tabular}
\end{center}
\end{table*}
Most recently, \cite{MUL14} found support for the presence of stars in Centaurus\,A's jet based on radio observations probing the subparsec scales. 

\cite{WYK13} proposed that a considerable fraction of the entrained baryonic material in Centaurus\,A consists of carbon, nitrogen and oxygen, and is passed to its giant (i.e. outer) lobes where it undergoes stochastic acceleration. A mixed composition of material that evolves into ultra-high energy cosmic rays (UHECRs) is consistent with recent results from large particle-detection instruments, which imply that the CR composition becomes heavier as a function of energy and that it may have more than one component (e.g. \citealp{ABR13, LET14, KAM14}).

The main objective of the present paper is to ascertain whether interactions with stellar winds could quantitatively be responsible for the observed X-ray emission in Centaurus\,A's jet and whether we can in principle account for the broad features of the spectrum of the present-day kpc-scale jet in this way. We then use the constraints on the stellar population that we have derived to calculate the mass-loss rates and nucleosynthetic isotope yields into the current and pre-existing jet to test the models and predictions of \cite{WYK13}. 

Section\,\ref{sect:params} introduces useful stellar and jet parameters within our basic model of jet-star interactions. The closest undertakings are probably the simulations by \cite{BOS12}, although they do not consider a stellar wind, the study of jet truncation via stellar winds by \cite{HUB06}, and the jet-massive star wind interactions modelled by \cite{ARA13}. The novel feature of our approach is the use of stellar evolution- and wind codes to carry out the modelling. In Section 3, we provide a resum\'e of the codes, and outline the restrictions and approximations. Section\,\ref{sect:results} presents the results of the modelling in terms of resultant synchrotron spectra, gives some estimates of mass-loss rates along with the impact of this material onto jet propagation, and discusses the likely baryon composition of the loaded kpc jet and lobes. The arguments are summarized and conclusions drawn in Section\,\ref{sect:summary}.

Throughout the paper, we define the energy spectral indices $\alpha$ in the sense $S_{\!\nu}\propto\nu^{-\alpha}$ and particle indices $p$ as $n(E)\propto E^{-p}$.

\section{Stellar and jet parameters} \label{sect:params}

\subsection{Basic model} \label{sect:model}

We consider the stand-off distance $R_0$, i.e. the distance from a star at which there is a balance between the momentum flux from the star and the momentum flux of the surrounding medium. The general approximation reads (e.g. \citealp{DYS75, WIL96})
\begin{equation}
R_0 = \Big(\frac{\dot{M} v_{\rm w}}{4\pi\rho\,v_*^2}\Big)^{1/2}\,. \label{eq:standgen}
\end{equation}
\begin{figure}
\center
\includegraphics[width=0.85\linewidth]{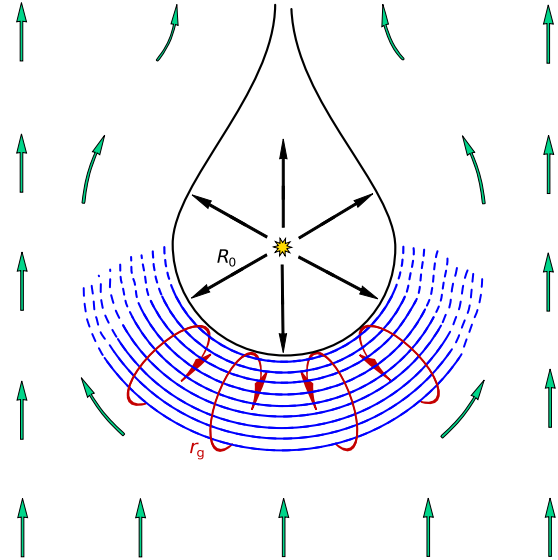}
\caption{Schematic representation of a star with wind in a jet flow, and the size and location of the shock and the shock region. Yellow dot represents the star, black arrows the stellar wind. The black solid line is the astropause, which happens at order of the stand-off distance $R_0$ from the star. Green arrows show the jet flow overall and around the astropause. The blue lines indicate the shocked region with $B$-field approximately perpendicular to the jet flow; the blue dashes to indicate uncertainty about how far out the shock region persists. The uppermost blue line can be thought of as the shock itself. The physical size of the shocked region scales with the stand-off distance. The red solid lines show high-energy electrons crossing and recrossing from the shocked to unshocked media.} \label{fig:fig2}
\end{figure}
Here $v_{\rm w}$ denotes the velocity of the isotropic\footnote{Stellar winds of so-called magnetic O and B stars (e.g. \citealp{BAB97}) are anisotropic, tunelled along the $B$-field axes. However, the fraction of such stars is small ($\sim7$ per cent; \citealp{GRU13}) and the anisotropy is, as a result of their rotation, expected to be largely washed out at $R_0$.} stellar wind (in the restframe of the star), $\rho$ the mass density of the surrounding medium and $v_*$ the relative space velocity of the star with respect to the surrounding fluid. For stars in a jet fluid, equation\,(\ref{eq:standgen}) becomes
\begin{equation}
R_0 = \Big(\frac{\dot{M} v_{\rm w}}{4\pi (U_{\rm j}/c^2)\,v_{\rm j}^2\,\Gamma_{\rm j}}\Big)^{1/2}\,,
\end{equation}
where $U_{\rm j}$ is the energy density of the jet and $\Gamma_{\rm j}=(1-\beta^2)^{-1/2}$ the jet Lorentz factor. We adopt $\beta=0.5c$ and $U_{\rm j}=8.77\times10^{-11}$\,erg\,cm$^{-3}$ \citep{WYK13}. $R_0$ essentially corresponds to the location of the contact discontinuity separating the shocked jet gas and the stellar wind.

We assume $v_{\rm w}= v_{\rm esc}$ and write
\begin{equation}
v_{\rm w}=v_{\rm esc} = \Big(\frac{2 G M_*}{R_*}\Big)^{1/2}\,,
\end{equation}
where $G$ is the gravitational constant, $M_*$ the stellar mass and $R_*$ the stellar radius.

We suppose that the thickness of the shocked region upstream of $R_0$ scales with $R_0$ (see Fig.\,\ref{fig:fig2}). Fermi\,I-type acceleration takes place as particles cross and recross the shock. Fermi\,I acceleration is no longer efficient when the gyroradius of a particle $r_{\rm g}$ exceeds that of the shocked region and so we can write
\begin{equation}
E_{\rm e,max} \simeq R_0\,e\,B\,, \label{eq:emax}
\end{equation}
with $E_{\rm e,max}$ the maximum electron energy, $e$ the electric charge and $B$ the magnitude of the magnetic field. 

For a jet dominated by leptons, the sound speed is higher than the adopted jet speed of $0.5c$, which means that it is not obvious that a strong bow shock will form in the jet material. However, $v_{\rm j}=0.5c$ is actually the projected jet speed: for the angle to the line of sight we are using, the internal speed in the jet would be faster than this. Moreover, it is reasonable to suppose that Centaurus\,A's jet has entrained enough material by the region we consider to have a lower internal sound speed.  

In the conditions of Centaurus\,A's jet, WR stars would have $R_0$ of order $16$\,pc, which corresponds, using the jet mean $B$-field strength\footnote{The $B$-field strength of a star falls off as the inverse of the distance squared from the star (magnetic dipole), and is thus smaller at $R_0$ than Centaurus\,A's jet mean $B$-field; this makes it unimportant for our calculations. More relevantly even, the shocked region is separated from the stellar wind material by a contact discontinuity, the astropause (Fig.\,\ref{eq:standgen}).} of $66$\,$\mu$G \citep{WYK13}, to $\gamma_{\rm e}\sim2\times10^{12}$ and $E_{\rm e,max}\sim1\times10^{18}$\,eV. For LBVs and normal O/B supergiants,\footnote{The qualification `normal O/B supergiants' encompasses O- and B-type stars on the main sequence, and later stages, until the onset of the LBV/WR phase.} we compute $R_0\sim3$\,pc and $E_{\rm e,max}\sim2\times10^{17}$\,eV, and for AGB stars $R_0\sim1$\,pc and $E_{\rm e,max}\sim7\times10^{16}$\,eV. In contrast, a typical M star only has $R_0\sim1\times10^{-5}$\,pc $\sim2$\,AU, and thus $\gamma_{\rm e}\sim1\times10^{6}$ and $E_{\rm e,max}\sim7\times10^{11}$\,eV in the jet. We have no evidence from observations of the jet for electrons above $\gamma_{\rm e}\sim1\times10^8$, that is $E_{\rm e,max}\sim5\times10^{13}$\,eV in the X-ray; the electron loss timescale there is tens to hundreds of years. The radiative loss limit where the synchrotron loss time equals the gyration time, sets a fundamental limit on the energies that the electron can reach: $\gamma_{\rm e,max} = (3\,e/\sigma_{\rm T}\,B)^{1/2}\sim 5.7\times10^9$ from which follows $E_{\rm e,max}\sim2.9\times10^{15}$\,eV (again for our mean $B$-field of $66$\,$\mu$G). Note that $E_{\rm e,max}$ for the high-mass-loss- and high-mass stars, and for the fundamental limit correspond to gamma-ray photons. 

The amount of jet energy intercepted by each star can be expressed as
\begin{equation}
E_{\rm intercept} \simeq \pi R_0^2\,U_{\rm j}\,v_{\rm j}\,\Gamma_{\rm j}\,. \label{eq:eintercept}
\end{equation}
Thus, for the ensemble of stars
\begin{equation}
E_{\rm intercept,all} \simeq \pi\,U_{\rm j}\,0.5 c\,\Gamma_{\rm j}\sum R_0^2\,. \label{eq:einterceptall}
\end{equation}
The total luminosity produced by the jet-star interaction cannot (greatly) exceed this value.

Finally, we can ask what spectrum is expected from the observations of the jet. To this end, we consider the injection electron distribution $i(E)$. We assume that for each star it is a power law with an injection index $p$ whose normalization scales with $E_{\rm intercept}$ and whose high-energy cutoff is given by equation\,(\ref{eq:emax}). $i(E)$ is an injection {\it rate} because it is the instantaneous spectrum produced by the acceleration. This leads to
\begin{equation}
\epsilon\,L_{\rm intercept} = \int_{E{\rm e,min}}^{E{\rm e,max}} E_{\rm e}\,i(E_{\rm e})\,$d$E_{\rm e}\,, \label{eq:lintercept}
\end{equation}
where $L_{\rm intercept}$ is the luminosity intercepted by the star and $\epsilon$ is an efficiency factor ($\epsilon<1$). For $p=2$ we have the simple result $\epsilon\,L_{\rm intercept}=i_0\,$ln$(E_{\rm e,max}/E_{\rm e,min})$, with $i_0$ the power-law normalization of the injection spectrum. We add the $i(E)$ up to obtain the total electron injection as a function of energy for all the stars: $I(E_{\rm e})=\sum i(E_{\rm e})$. We calculate the synchrotron emissivity from the jet, considering $n_{\rm e}(E_{\rm e})$, the number of electrons in the jet as a function of electron energy. $n_{\rm e}(E_{\rm e})$ obeys d$n_{\rm e}(E_{\rm e})/$d$t = I(E_{\rm e}) - n_{\rm e}(E_{\rm e})/t_{\rm esc} - n_{\rm e}(E_{\rm e})/t_{\rm loss}(E_{\rm e})$. Here $I(E_{\rm e})=\sum i(E_{\rm e})$, $t_{\rm esc}$ is the (energy-independent) time for material to move out of the jet ($\sim l_{\rm j}/2v_{\rm j}$, with $l_{\rm j}$ the length of the jet, divided by two to get the typical distance an electron has to travel to escape the jet) and $t_{\rm loss}(E_{\rm e})$ is the energy-dependent electron loss timescale, whereby $t_{\rm loss}$ goes as $1/E_{\rm e}$. For a steady-state jet we set d$n_{\rm e}(E_{\rm e})/$d$t=0$ for all energies. Then
\begin{equation}
n_{\rm e}(E_{\rm e}) = \frac{I(E_{\rm e})}{1/t_{\rm esc}+1/t_{\rm loss}}\,. \label{eq:neee}
\end{equation}
The synchrotron emissivity can then be calculated from $n_{\rm e}(E_{\rm e})$ in the standard way (e.g. \citealp{RYB86}).

\subsection{Census of old- and young-population stars} \label{sect:census}

\subsubsection{Distribution, ages and metallicities of old stars} \label{sect:distr_old}

Earlier works (e.g. \citealp{SOR96}) have provided evidence for the existence of hundreds of red giant branch (RGB) stars and AGB stars in the halo of NGC\,5128, and have hinted at more than one epoch of star formation on Gyr timescales. \cite{REJ11} have argued for two old stellar populations throughout NGC\,5128: $70-80$ per cent of stars forming older population with ages of $12\pm1$\,Gyr and with metallicities consistent with values $Z=0.0001- 0.04$, while $20-30$ per cent stars have an age in the range $2-4$\,Gyr with a minimum metallicity of $0.1$ to $0.25$ the solar value $Z_{\odot}$ ($Z_{\odot}=0.0198$). Given that the majority of the stars are fairly old and that the overall metallicity distribution function peaks close to log\,$(Z/Z_{\odot})=-0.3$ (\citealp{REJ11}, their fig.\,1), we adopt $75$ per cent of old stars of $12$\,Gyr at $Z=0.004$ and $25$ per cent of old stars of $3$\,Gyr at $Z=0.008$ as the `average' ages and metallicities for our modelling (see Table\,\ref{tab:para2}).

\subsubsection{Distribution, ages and metallicities of young stars} \label{sect:distr_young}

Centaurus\,A's optical and infrared emission shows a pronounced dust lane (e.g. \citealp{DUF79, ECK90}), with a starburst of $\sim60$\,Myr \citep{UNG00} which is plausibly a result of a Large Magellanic Cloud-type galaxy (with no black hole) merging with NGC\,5128, as suggested by the amount of molecular and dust material (F. Israel, private communication). The dust lane hinders optical to ultraviolet studies of the inner $\sim1$\,kpc (projected) of the jet.

If the starburst in Centaurus\,A formed stars only once, about $60$\,Myr ago, stars with an initial mass $M_{\rm init}\ga6$\,M$_{\odot}$ may no longer be present in it. The supernova SN\,1986G that occured in the south-east part of the starburst \citep{EVA86, CRI92}, well away from the jet, does not provide evidence for a current high-mass star presence in the starburst, being of Ia type. However, there is probably intermittent star formation activity since the (most recent) merger and so we expect O and B stars to form. \cite{MOL81} and \cite{MIN04} have reported blue star clusters in parts of the starburst and WR-type emission based on Very Large Telescope (VLT) observations, and blue star clusters are also directly visible in {\it Hubble Space Telescope} ({\it HST}) images \citep{VIL05}.

Examples of star-forming regions not coincident with the dust lane include young stars in a number of filaments along the jet and beyond its radial extent (e.g. \citealp{GRA98, REJ01, REJ02, OOS05, CRO12}); some of this star formation could be triggered (directly or indirectly) by the (current or pre-existing) jet activity (for example, \citealp{GAI12} have shown, via numerical simulations, that jets can trigger star formation during the initial phases of their expansion), or perhaps by a starburst wind originating in star associations embedded in the dust lane. Where estimated, the ages of these stars are in the range $\sim1-15$\,Myr \citep{FAS00, GRA02, REJ01, REJ02, CRO12}. Young stars may be present also elsewhere in the NGC\,5128 field, counting in the volume of the current jet.

On the basis of metallicities obtained from spectroscopy of H\,{\small II} regions of the starburst \citep{MOL81, MIN04} and from {\it Suzaku} X-ray line observations of diffuse plasma of the circumnuclear material \citep{MAR07}, which are both close to solar, we adopt $Z=0.02$ for the young stellar component (Table\,\ref{tab:para2}).

\subsubsection{Number of stars in the jet} \label{sect:numberstars}

A convenient approach for assessing the number of stars in the jet volume is to determine the observed luminosity ($L_{\rm obs}$) in physical units from aperture photometry, work out the normalization factor from the SSE-synthesised (see Section\,\ref{sect:code}) stellar population and assume some jet geometry.
\begin{table}
\begin{center}
\caption{Adopted stellar populations ages, metallicities $Z$ and numerical fractions $f_*$.}
\label{tab:para2}
\begin{tabular}{ccc}
\hline \\ [-1.5ex]
Age & $Z$ & $f_*$ \\
(Myr) &  & (per cent) \\
[-1.5ex] 
\multicolumn{3}{l}{} \\
\hline \\ [-1.5ex]
$12\,000$ & $0.004$ & $75$ \\
$3\,000$  & $0.008$ & $25$ \\
$0-60$    & $0.02$  & $0-2$ \\
[1ex] \hline
\end{tabular} \\
\end{center}
\end{table}

To determine $L_{\rm obs}$ from aperture photometry, we have used the $R$-band photometry\footnote{$R$-band is less affected by obscuration than $B$-band.}$^,$\footnote{{\tt http://leda.univ-lyon1.fr/fG.cgi?n=3\&o=NGC5128}} for the decimal logarithm of the diameter of the aperture log$(A) = 2.02$ yielding $5.2$\,arcmin (i.e. $5.7$\,kpc projected size, close to the adopted projected jet length of $4.5$\,kpc, see also Section\,\ref{sect:code}). We synthesise, utilizing the SSE routine (described in Section\,\ref{sect:code}), a population of $N$ stars (with the stellar parameters summarized in Table\,\ref{tab:para2}) and for each star compute the luminosity at the reference frequency ($L_{\rm ref}$), assuming that the star is a black body with radius $R = R_{\rm eff}$ and temperature $T = T_{\rm eff}$. We then have $L_{\rm ref} = 2.84\times10^{-20}\,10^{-7.58/2.5} 4 \pi (3.8\times10^6 c)^2$, where the numerical factor (in erg\,s$^{-1}$\,cm$^{-2}$\,Hz$^{-1}$) is the flux density equivalent to the zero of magnitude in the $R$-band \citep{ZOM07}. We next add up these luminosities to obtain their $L_{\rm tot}$. The total number of stars required is then $N_{\rm tot}=N\,L_{\rm obs} / L_{\rm tot}$. This yields a total number of stars in the jet of $\sim8\times10^8$. The aperture photometry luminosity should be corrected for the fact that the region we are seeing is not a sphere of radius $R$ but the integral over a cylindrical slice through the galaxy of radius $R$. However, that correction will probably be of order unity, and in the opposite sense to the correction for the dust lane, which we can not do.

We have very little information to allow us to estimate the fraction of young stars, hence we ought to ask what fraction of young stars is reasonable. A starburst lasting $60$\,Myr with a star formation rate (SFR) of $1-2$\,M$_{\odot}$\,yr$^{-1}$ (which is only the same as the Milky Way's, see \citealp{ROB10} and references therein) produces $6-12\times10^7$\,M$_{\odot}$ of young stars. This would yield around $1$ per cent young stars. The SFR estimated from the far infrared luminosity of Centaurus\,A's starburst of $L_{\rm FIR}=5.8\times10^9$\,L$_{\odot}$ at $3$\,Mpc (\citealp{ECK90}; i.e. $L_{\rm FIR}=3.6\times10^{43}$\,erg\,s$^{-1}$ at $3.8$\,Mpc) multiplied by the factor $4.5\times10^{-44}$ (see \citealp{KEN98}, their relation for $\la100$\,Myr old starbursts), gives a SFR of $\sim1.6$\,M$_{\odot}$\,yr$^{-1}$, consistent with the above value. Note that if the starburst is confined to the dust lane, it is possibly toroidal, i.e. not filled in the centre, with the result that the jet is traversing the starburst for only a fraction of its obscured part, if at all. That would bring the young stellar content close to zero.\footnote{Additional support for a low fraction of young stars in the current jet comes from (publicly available) {\it GALEX} images that seem to show that young stars are very present but only in the dark lane. Since this dark lane genuinely is an extended thin disc (ETD) that only exists at relatively large radial distances from the core \citep{NIC92}, these stars will not be noticed by the jet. Moreover, the analysis of the circumnuclear disc (CND) by Israel et al. (in preparation) suggests that there are no young stars and present-day star formation within a few hundred pc from the core.} These uncertainties mean that it is most appropriate to ask about the viability of the model as a fraction of young stars, modelling the spectrum with a range of plausible values. We chose to run the codes with $0, 0.1$, $0.2$, $0.3$, $0.5$, $1$ and $2$ per cent young stars (Table\,\ref{tab:para2}, and see Sections\,\ref{sect:eintercept} -- \ref{sect:spectrum_results}).

\subsection{Broad-band spectrum of the jet} \label{sect:spectrum_intro}

Observationally, the best constrained broad-band spectrum, radio through X-ray, of the kpc jet is presented in \cite{HAR06}; as in FR\,I jets in general (e.g. \citealp{HAR02}), it is inconsistent with a single power-law model. \cite{HAR06} define three sections outside the dust lane (necessary to get the IR and optical data points): the `inner region' (see Fig.\,\ref{fig:fig1}; not to be confused with the term `inner jet' which includes the entire part upstream of $2.4$\,kpc as well), and further downstream the `middle region' and `outer region'.

\subsubsection{X-ray component} \label{sect:spectrum_x}

\cite{HAR11} have measured Centaurus\,A's jet X-ray diffuse luminosity from {\it Chandra} data of $\sim134$\,nJy which translates, adopting the distance to Centaurus\,A of $3.8$\,Mpc, to $L_{\rm X}\sim6\times10^{38}$\,erg\,s$^{-1}$. For all of the knots, we adopt the sum of the X-ray luminosities from {\it Chandra} observations by \cite{GOO10} that gives $\sim126$\,nJy, i.e. $L_{\rm X}\sim5\times10^{38}$\,erg\,s$^{-1}$ at the $3.8$\,Mpc distance, and thus $L_{\rm X}\sim1.1\times10^{39}$\,erg\,s$^{-1}$ for the combined diffuse- and knot emission. 

Most stars are intrinsic X-ray sources. Taking all $8\times10^8$ stars in the jet to be solar-like (the mean steady X-ray luminosity of the Sun is $L_{\rm X,\odot}\sim1\times10^{27}$\,erg\,s$^{-1}$), we would have a total X-ray luminosity $L_{\rm X}\sim8\times10^{35}$\,erg\,s$^{-1}$ from such stars. This is three orders of magnitude below the measured X-ray luminosity of the diffuse emission of the jet and thus negligible. The high $L_{\rm X}$ values of individual high-mass-loss stars (Table\,\ref{tab:para}) are also well below what is observed for the X-ray knots. X-ray binaries (Low-Mass X-ray Binaries, LMXBs, and High-Mass X-ray Binaries, HMXBs) have a few orders of magnitude higher $L_{\rm X}$ than the single-evolution stars considered above, typically in the range $L_{\rm X}\sim1\times10^{35}$\,erg\,s$^{-1}$ to $L_{\rm X}\sim1\times10^{38}$\,erg\,s$^{-1}$. \cite{GOO10} have found no evidence for the X-ray-bright knots in Centaurus\,A's jet being associated with LMXBs.

\section{Framework and approximations} \label{sect:code}

We used the stellar evolution code described by \cite{HUR00}, and stellar wind codes by \cite{CRA11} and \cite{VIN99, VIN00, VIN01}, either translated into Python or with Python interfaces generated by us. We wrote additional Python codes to extract, from the elemental codes, the luminosities and mass-loss rates for stellar populations with our adopted age and metallicity constraints, and to compute the parameters $E_{\rm e,max}$, $E_{\rm intercept,all}$, energy intercepted by stars producing X-rays $E_{\rm intercept,X}$, number of stars with spectrum reaching frequencies above $10^{16.5}$\,Hz, the total luminosity $L_{\rm tot}$ and the total mass-loss rate $\dot{M}$.

The Single-Star Evolution (SSE) routine\footnote{\tt http://astronomy.swin.edu.au/$\sim$jhurley} \citep{HUR00}, based on a number of interpolation formulae as a function of $M_{\rm init}$, stellar age and $Z$, provides predictions for $\dot{M}$ for phases with high mass-loss rates. To fill in for the missing mass-loss rates, we added the BOREAS routine\footnote{\tt http://www.cfa.harvard.edu/$\sim$scranmer} \citep{CRA11}, which computes $\dot{M}$ for cool main-sequence stars and evolved giants, up to $T_{\rm eff}=8000$\,K, and the mass-loss prescription\footnote{\tt http://star.arm.ac.uk/$\sim$jsv/Mdot.pro} of \cite{VIN99, VIN00, VIN01} on the basis of Monte Carlo radiative transfer calculations for high-mass stars, which is valid in the range $8000\le T_{\rm eff}\le50\,000$\,K.

The calculation of $R_0$, by calling the SSE routine for the stellar masses and radii, inherently involves $v_{\rm esc}$ for calculating the wind speed, and hence assumes $v_{\infty}/v_{\rm esc}=1$, which is rough (see in this context Table\,\ref{tab:para}, and also e.g. \citealp{JUD92} for low-mass stars). The jet geometry adopts a projected length of $4.5$\,kpc which translates to a physical length of the jet of $5.87$\,kpc at the viewing angle $50^{\circ}$, and we have treated the jet as a simple cone with an opening angle $15^{\circ}$. For the initial mass function (IMF), i.e., the distribution of stellar masses at formation, we adopt the Salpeter IMF \citep{SAL55}: $\phi(M_{\rm init})=$ d$N(M_{\rm init})\propto M_{\rm init}^{-x}\,$d$M_{\rm init}$, with $x=2.35$ between $0.5$ and $120$\,M$_{\odot}$ and a `heavy' $x=1.3$ between $0.08$ and $0.5$\,M$_{\odot}$ \citep{REJ11}. We next draw stars from the IMF, using the cumulative probability distribution, and calculate their present-day mass and radius using SSE, their mass-loss rate as outlined above, and the maximum electron energy of particles accelerated in stand-off shocks as described in equations\,($1-4$). A normalization factor is included to account for the number of real ($N_{\rm real}$) and simulated ($N_{\rm simulated}$) stars. We count stars whose synchrotron emission would extend above $10^{16.5}$\,Hz in order to separately trace the effect of young massive stars plus AGB stars.

We neglect the effect of windless stellar objects such as white dwarfs (WDs): the effective cross section for interaction with compact objects is completely negligible.\footnote{A M-type star with $R_0$ of order $1$\,AU intercepts about a factor $10^9$ more power than a windless WD.} We presume that the starburst has a uniform distribution through the centre of the galaxy. We assume that all stars in the jet are field stars, not in clusters,\footnote{The cluster formation efficiency (CFE) is generally high, possibly up to $50$ per cent, shortly after the stars have formed (e.g. \citealp{KRU12}). On a Gyr timescale, the fraction of stars in clusters is found to be much smaller than this, for NGC\,5128 in particular only of order $0.1-0.3$ per cent \citep{HARR02}. However, the CFE in the dust lane may still be relatively high.\label{foot:cfe}} therefore also neglecting a possible occurence of multiple stars at close distance, intercepting the same arc of the jet and hence reducing $E_{\rm intercept,all}$. We do not account for binary interaction effects on the mass-loss rate. We disregard second-order effects such as the extent to what the various type of stars are affected by the jet plasma, potentially leading to changes in $R_0$ and the mass-loss rate. The chance that a supernova exploded within the jet boundary in the lifespan of $\sim2$\,Myr is negligible (and there is thus far any observational evidence for a supernova remnant inside the present-day jet, see also Section\,\ref{sect:introduction}), hence we do not account for this either. Since our modelling is designed for a mean population of stars at any given time in the jet, orbital star crossing does not affect our treatment. (It is the synchrotron timescale that matters, which is tens to thousands of years. As long as the evolutionary timescale is larger than or comparable to that, we are justified in assuming that the stars do not change.)

The final computations, in which we used $100$\,million simulated stars to avoid small-number effects in the stars that produce X-rays, were done on the University of Hertfordshire cluster.\footnote{\tt http://stri-cluster.herts.ac.uk/}

\section{Results and interpretation} \label{sect:results}

\subsection{$\bf {\it E}_{\rm \bf intercept,X}$ criterion} \label{sect:eintercept}

First, we investigate whether we can meet the $E_{\rm intercept,X}$ criterion, i.e. whether the stars that are supposed to produce the X-rays intercept enough energy to allow them to do so, for a plausible fraction of young stars.
\begin{table}
\begin{center}
\caption{Parameters obtained from modelling, for various fractions of young (i.e. $0-60$\,Myr) stars $f_*,{\rm young}$ and with $N_{\rm simulated}=1\times10^8$ stars: energy intercepted by stars leading to X-ray emission $E_{\rm intercept,X}$, number of stars leading to X-ray emission $N_{\rm X}$, energy intercepted by all stars $E_{\rm intercept,all}$ and entrainment rate $\Psi$.}
\label{tab:para3}
\begin{tabular}{lcccc}
\hline \\ [-1.5ex]
$f_*,{\rm young}$ & $E_{\rm intercept,X}$ & $N_{\rm X}$ & $E_{\rm intercept,all}$ & $\Psi$\\
(per cent) & (erg\,s$^{-1}$) & & (erg\,s$^{-1}$) & (M$_{\odot}$\,yr$^{-1}$) \\
[-1.5ex] 
\multicolumn{5}{l}{} \\
\hline \\ [-1.5ex]
$0$   & $9.7\times10^{39}$ & $1.73\times10^4$ & $1.8\times10^{40}$ & $2.3\times10^{-3}$ \\
$0.1$ & $6.2\times10^{40}$ & $1.84\times10^4$ & $7.1\times10^{40}$ & $2.9\times10^{-3}$ \\
$0.2$ & $4.9\times10^{40}$ & $2.0\times10^4$  & $5.8\times10^{40}$ & $3.0\times10^{-3}$ \\
$0.3$ & $3.5\times10^{40}$ & $2.16\times10^4$ & $4.4\times10^{40}$ & $4.4\times10^{-3}$ \\
$0.5$ & $1.0\times10^{41}$ & $2.47\times10^4$ & $1.0\times10^{41}$ & $4.0\times10^{-3}$ \\
$1$   & $3.7\times10^{41}$ & $3.40\times10^4$ & $3.8\times10^{41}$ & $7.4\times10^{-3}$ \\
$2$   & $5.1\times10^{41}$ & $5.04\times10^4$ & $5.2\times10^{41}$ & $1.5\times10^{-2}$ \\
[1ex] \hline
\end{tabular} \\
\end{center}
\end{table}

Table\,\ref{tab:para3} shows the output from our modelling, revealing  $E_{\rm intercept,X}$ between $9.7\times10^{39}$ and $5.1\times10^{41}$\,erg\,s$^{-1}$, depending on the fraction of young stars. There is some scatter in the results, as expected: if there are potentially very young massive stars in the jet then they will have a significant effect on $E_{\rm intercept,X}$ and $E_{\rm intercept,all}$. Even with a zero fraction of young stars, we do not run into difficulties in producing the observationally determined $L_{\rm X}\sim1\times10^{39}$\,erg\,s$^{-1}$ (see Section\,\ref{sect:spectrum_x}). 
\begin{figure*}
\includegraphics[width=1.0\textwidth]{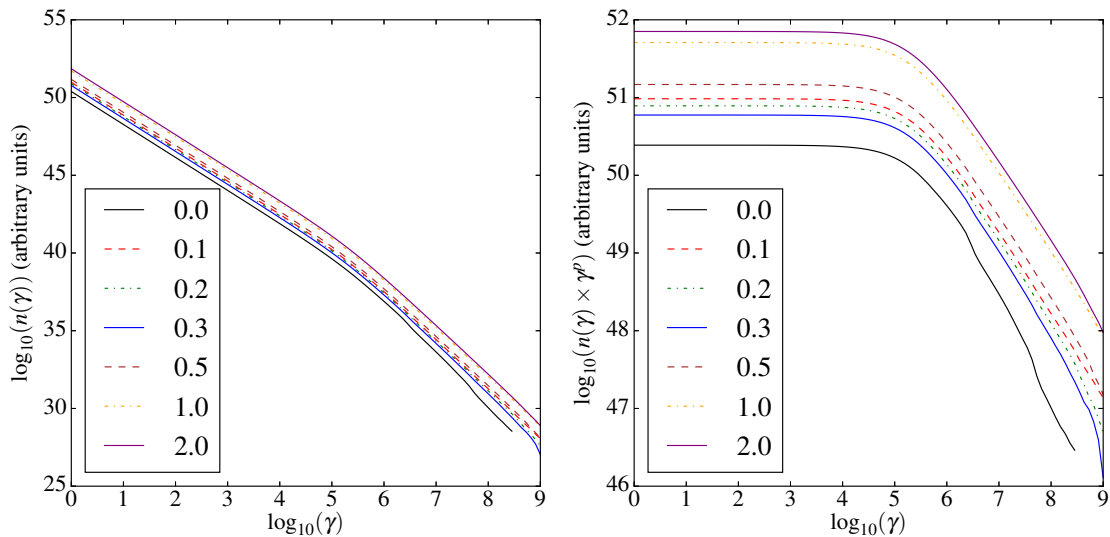}
\caption{Electron distributions of the kpc-scale jet corresponding to the synchrotron spectra of Fig.\,\ref{fig:fig4}, with old and young ($0, 0.1, 0.2, 0.3, 0.5, 1$ and $2$ per cent) stellar components. The left-hand panel plots $n_{\rm e}(\gamma)$ as a function of $\gamma$. The right-hand panel plots $n_{\rm e}(\gamma)\gamma^2$ to demonstrate the near $\gamma^{-2}$ power law up to $\gamma\sim10^4$, and the steepening of the density spectrum at higher $\gamma$. These results are for a run with $N_{\rm real}=8\times10^8$ and $N_{\rm simulated}=1\times10^8$ stars.} 
\label{fig:fig3}
\end{figure*}

Thus, our model is energetically capable of producing the observed X-ray emission. There are likely enough high-mass-loss stars and normal O/B supergiants to provide all the discrete X-ray knots and produce the diffuse emission. This alleviates the need for additional synchrotron-producing mechanisms in Centaurus\,A's jet such as stochastic acceleration, shear or magnetic reconnection, although we do not exclude these processes making additional contribution in parts of the jet. Our work assumes $100$ per cent acceleration efficiency, which may be too optimistic; at the other hand, we use the mean jet $B$-field to estimate maximum energies, while, presumably, the post-shock field is amplified. We have ignored shielding of stars by one another, which may in fact occur to some degree for the young component (see footnote\,\ref{foot:cfe}) and so reduce $E_{\rm intercept,X}$ and $E_{\rm intercept,all}$. Another caveat is that the viewing angle is not extremely well known; this could affect the number of jet-contained stars and hence impact on $E_{\rm intercept,X}$, $E_{\rm intercept,all}$ and the resulting synchrotron spectrum. Given that the opening angle of Centaurus\,A's jet is $\sim12^{\circ}$ on subparsec and parsec scales (e.g. \citealp{HOR06, MUL11, MUL14}), our assumption of the jet opening angle of $15^{\circ}$ overall means a slight overestimate of the number of stars in the jet. However, none of these effects, with the exception of a particle-acceleration efficiency $\ll1$, affect our basic conclusion. Our analysis of the spectrum (and the implications for the fraction of young stars) depends completely on equation (\ref{eq:emax}). This rests only on the statement that Fermi acceleration becomes inefficient when the electron gyroradius becomes larger than the scale of the system, and on the assumption that the shocked region size scales linearly with $R_0$. 

An additional statement we can make based on the zero-fraction young stars outcome is that there must be at least $1\times10^4$ AGB stars in the jet. We make use of this value in Section\,\ref{sect:enrichment_quantitative}, below.

\subsection{Electron spectrum} \label{sect:espectrum_results}

In a subset of our codes, we implement the calculation of $n_{\rm e}(\gamma)$, using equation\,(\ref{eq:neee}). We compute $i(\gamma)$ for each star and add them up to obtain $I(\gamma)$, using a modified version from the above, scaled using equation\,(\ref{eq:neee}), to get the actual electron energy spectrum in the jet $n_{\rm e}(\gamma)$. We include a scaling with log$(\gamma_{\rm max})$ from equation\,(\ref{eq:lintercept}). The results are shown in Fig.\,\ref{fig:fig3}: the left-hand panel is with $n_{\rm e}(\gamma)$ and the right-hand panel with $n_{\rm e}(\gamma)\gamma^2$ to emphasise the steepening at high energies. Note that $n_{\rm e}(\gamma)$ steepens to a power-law index of greater than $p=3$ ($p\sim3.3$ at high $\gamma$), consistent with observations. Moreover, the onset of steepening is at the appropriate frequency.
\begin{figure*}
\includegraphics[width=\linewidth]{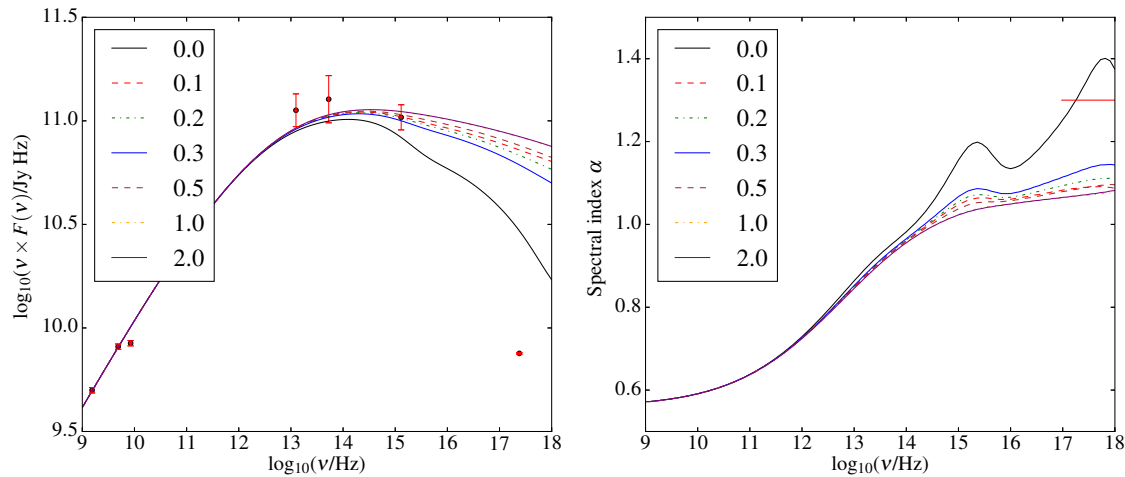}
\caption{Left: Broad-band spectrum of the kpc-scale jet with old and young ($0, 0.1, 0.2, 0.3, 0.5, 1$ and $2$ per cent) stellar components. A run with $N_{\rm real}=8\times10^8$ and $N_{\rm simulated}=1\times10^8$ stars. Radio through to X-ray data points (red, filled circles) of the `inner region' of the jet (corresponding to the jet on $2.4-3.6$\,kpc scale projected) are adopted from \citet{HAR06}. The points represent diffuse- and knot emission. Vertical bars are $1\,\sigma$ errors on the data; where omitted, the error bars are smaller than the symbols. Right: Spectral index of the kpc-scale jet as a function of frequency, with the old and young stellar components as above. The red horizontal line indicates the mean X-ray spectral index.} 
\label{fig:fig4}
\end{figure*}

We ascribe the high-$\gamma$ tail to the effect of either of AGB, LBV and WR stars and normal O/B supergiants (which have $R_0$ comparable to LBVs; see Section\,\ref{sect:model}). However, we have no means of discriminating between the above stellar types in the way they affect the high-$\gamma$ tail.  
  
An additional point to make is that the stellar wind density $n_{\rm w}$, particularly for the high-mass-loss stars, is relatively high (see Table\,\ref{tab:para}), even at $R_0$, which could lead to a production of secondary electrons through hadron-hadron collisions. This affects our calculation in so far as it would mean that the injection spectrum would not be a single power law, but it does not affect the energetic constraints on the X-ray emission. However, it is doubtful whether the stellar wind material of a particular star will actually enter the shocked region upstream of it. More plausible is perhaps that most of the material is ablated and entrained for encounters downstream, lowering somewhat the probability of the secondary electron production.

\subsection{Radio to X-ray spectrum} \label{sect:spectrum_results}

The resultant, broad-band synchrotron spectrum of the kpc jet (Fig.\,\ref{fig:fig4}) shows different cutoffs, inconsistent with a single power-law. Our fine-tuning exercise on the fraction of young stars in the jet reveals $0-0.5$ per cent young stars.

The left-hand panel of Fig.\,\ref{fig:fig4} displays the broad-band spectrum of the kpc jet with the observational data points for the `inner region' of the jet, i.e. $2.4-3.6$\,kpc (projected) from the galaxy centre (see Fig.\,\ref{fig:fig1}), as in \cite{HAR06}. Here we have converted to $\nu F_{\nu}$. The broad features of Fig.\,\ref{fig:fig4} demonstrate that all fractions of young star values well reproduce the spectrum up to the optical. The X-ray and higher frequencies are very sensitive to the fraction of young stars. The normalization is adjusted to make the curve go through the radio data point, which is equivalent to assuming that all the electrons in the jet are produced by stellar interactions; factors of order unity in the assumptions from Section\,\ref{sect:model} could make a significant difference to details of the synchrotron spectrum here.

In our fit, the `inner region' has the same radio to X-ray ratio as the whole $4.5$\,kpc jet. Note that this is arbitrarily normalized to the radio fluxes, rather than being directly normalized from $E_{\rm intercept,all}$. We do not assert that the proposed acceleration mechanism might produce all the radio emission. It is possible that the emission at bands other than the X-ray (partially) originates from electrons that have not been accelerated {\it in situ}. For example, since we measure radio emission there, some electrons must be accelerated in the pc-scale jet and these can be transported to kpc scales. If some of the particle acceleration that gives the radio/optical emission is not related to stellar interactions, which is conceivable, then the normalization of all the curves might go down. 

The right-hand plot in Fig.\,\ref{fig:fig4} shows the spectral index variation with frequency. This demonstrates that we can reproduce the mean X-ray spectral index ($\alpha\sim1.3$; red horizontal line) for sensible fractions of young stars of order $0-0.5$ per cent. Moreover, it illustrates that if there are no young stars we obtain very steep spectral indices; this is a possible explanation for the very steep spectral index ($\alpha\sim2.5$) of the jet seen at large distances from the core by \cite{HAR07}.

In summary, we reproduce the broad-band spectrum of the present-day jet up to optical frequencies. For fractions of young stars consistent with our expectations, we are within a factor $2-3$ of the X-ray flux, which we regard as a good outcome given the sensitivity of the X-ray spectrum to the details of assumptions in our models, and the fact that some fraction of the low-energy electron population is probably accelerated elsewhere.

\subsection{Entrainment rates} \label{sect:rates}

We next re-examine the entrainment rates within the jet boundaries through adding the mass-loss rates for our stellar populations and their normalization and scaling them to obtain the total internal entrainment rate over the jet.

We compute a lower limit to the entrainment rate of $\Psi\sim2.3\times10^{-3}$\,M$_{\odot}$\,yr$^{-1}$. Compared to the result of \cite{WYK13}, who calculated $\Psi\sim6.8\times10^{22}$\,g\,s$^{-1}$ (i.e. $\sim1.1\times10^{-3}$\,M$_{\odot}$\,yr$^{-1}$), with older stars as the major contributors, this is higher by a factor $\sim2$, which is in good agreement, given the simple assumptions in the earlier work. Our result implies internal entrainment of $4.6\times10^3$\,M$_{\odot}$ during the lifetime of the current jet, and of $1.3\times10^6$\,M$_{\odot}$ over the lifetime of the giant lobes. Including the external entrainment contribution, i.e. material picked up from the galaxy ISM and transported downstream the jet, estimated by \cite{WYK13} as $\Psi\sim3.0\times10^{21}$\,g\,s$^{-1}$ (i.e. $\sim4.7\times10^{-5}$\,M$_{\odot}$\,yr$^{-1}$), will not markedly increase the above figures for the jet and lobes.

Before we assess the entrainment rates in individual nuclei (Section\,\ref{sect:enrichment}), we briefly address the impact of the entrainment on the jet flow speed.

\subsection{Jet deceleration} \label{sect:deceleration}

We could ask what speed does the current jet have, if it is initially baryon-free with a speed around $0.5c$, and assuming that momentum is conserved. The relativistic leptonic jet fluid behaves as having a density $U_{\rm j}/c^2$. Hence, if the relativistic momentum flux is $\dot M v\,\Gamma$, we have a condition
\begin{equation}
\pi\,r_{\rm j}^2\,\frac{U_{\rm j}}{c^2}\,v_{\rm j,1}^2\,\Gamma_{\rm j,1} = \dot{M}v_{\rm j,2}\,\Gamma_{\rm j,2}\,,
\end{equation}
where we assume that the momentum at the end of the jet is dominated by entrained material. Taking $P_{\rm j}=1\times10^{43}$\,erg\,s$^{-1}$ (Section\,\ref{sect:introduction}), $\Gamma_{\rm j,1}\sim1.15$ (Section\,\ref{sect:model}) and the internal entrainment rate of $\Psi\sim1.4\times10^{23}$\,g\,s$^{-1}$ (i.e. $\sim2.3\times10^{-3}$\,M$_{\odot}$\,yr$^{-1}$; see the foregoing section), we solve for $v_{\rm j,2}\Gamma_{\rm j,2}$ to find $0.04c$. The jet may gain some momentum due to a possible external pressure gradient, but it is not expected to increase the derived value of $0.04c$ by much. Thus, the material expected to be entrained via stellar winds can lead to a significant slow-down of the current jet, without completely disrupting it. 

The higher entrainment rates for a larger fraction of young stars in the jet (Table\,\ref{tab:para3}) suggests that the jet would decelerate very quickly if there were a high fraction of young stars. Given this together with the X-ray normalization and spectral index results, and the fact that there is no energetic reason to prefer high fraction of young stars, low values for the fraction of young stars are favoured. However, note that the fraction of young stars is integrated over the jet, and may in reality vary spatially (see Section\,\ref{sect:distr_young} for discussion of the nuclear starburst and how far out it extends).

\subsection{Enrichment in nuclei} \label{sect:enrichment}

In this section, we present nucleosynthetic yields of the most abundant isotopes in stellar winds for stellar populations in the (current and pre-existing) jet, given the age and metallicity constraints.

\subsubsection{General picture} \label{sect:enrichment_general}

Winds of low-mass ($M_{\rm init}\la1$\,M$_{\odot}$) main-sequence stars are very weak and predicted to contain isotopes only at the initial abundances; to first order, the composition of higher-mass main-sequence winds is also the initial one. This ensures that $\chem \iso{4}{}He$, $\chem \iso{16}{}O$, $\chem \iso{12}{}C$ and $\chem \iso{14}{}N$ are the most abundant isotopes in their winds, albeit at minute quantities.\footnote{Protons, while not a product of stellar nucleosynthesis, are still the most abundant component of stellar winds.} AGB stars, the most numerous ingredient among our high-mass-loss star sample, shed about $1-7$\,M$_{\odot}$ of material per star into their surroundings during the lifetime of the AGB phase (see Table\,\ref{tab:para}). Stellar nucleosynthesis of AGB stars and their nucleosynthetic yields have recently been reviewed by Karakas \& Lattanzio (2014; see also \citealp{HER05, CRI09}); here we briefly highlight the relevant features.

The lowest-mass AGB stars ($M_{\rm init}\sim0.8-1$\,M$_{\odot}$) that evolve in under $12$\,Gyr expel solely the products of $\chem \iso{}{}H$-burning which are mixed to the surface by the first dredge-up. The net yields are dominated by $\chem \iso{3}{}He$, $\chem \iso{4}{}He$, $\chem \iso{13}{}C$ and $\chem \iso{14}{}N$. Stars with masses between about $1.5$ and $3-4$\,M$_{\odot}$, depending on metallicity, expel the products of (partial) $\chem \iso{}{}He$-burning along with $\chem \iso{}{}H$-burning; hence, the net yields are dominated by $\chem \iso{4}{}He$, $\chem \iso{12}{}C$, $\chem \iso{14}{}N$ and $\chem \iso{22}{}Ne$.\footnote{Lower-mass AGB stars show $\chem \iso{16}{}O$-condensations in their envelopes and are therefore generally referred to as oxygen-rich AGB (e.g. \citealp{SPE00, ATH02}), despite the absence of nucleosynthetic $\chem \iso{16}{}O$ in their winds.}  The most massive AGB stars, including the so-called super-AGB stars, experience hydrogen burning at the base of the convective envelope and the net yields are once again dominated by $\chem \iso{}{}H$-burning but with an added primary component from the dredge-up of He-shell products to the $\chem \iso{}{}H$-burning region. The net yields of stars in this upper-mass range $4-10$\,M$_{\odot}$ are dominated by $\chem \iso{4}{}He$, $\chem \iso{14}{}N$, $\chem \iso{23}{}Na$, $\chem \iso{25}{}Mg$ and $\chem \iso{26}{}Mg$ (e.g. \citealp{SIE10, DOH14}). In all cases, the dominant element expelled by a population of low- and intermediate-mass stars is still hydrogen, but with enhanced levels relative to the initial abundances of the isotopes mentioned above.

Pre-supernova phases of $M_{\rm init}\ga10-12$\,M$_{\odot}$ stars copiously produce $\chem \iso{4}{}He$, $\chem \iso{12}{}C$, $\chem \iso{14}{}N$ and $\chem \iso{16}{}O$ (e.g. \citealp{HIR05, KOB06, CHI13}). At somewhat lower levels -- and this is relevant to AGB stars as well as massive stars -- nuclei such as $\chem \iso{3}{}He$, $\chem \iso{7}{}Li$, $\chem \iso{13}{}C$, $\chem \iso{17}{}O$, $\chem \iso{18}{}O$, $\chem \iso{20}{}Ne$, $\chem \iso{24}{}Mg$, $\chem \iso{28}{}Si$, $\chem \iso{32}{}S$, and the radioactive $\chem \iso{26}{}Al$ and $\chem \iso{60}{}Fe$ (with half-lives, respectively, $\sim0.7$ and $2.6$\,Myr) are also expected. The stable nucleus $\chem \iso{56}{}Fe$, often considered in contemporary studies of the composition of (U)HECRs by authors engaged in particle detection experiments, is only formed during supernova explosions through the decay channel $\chem \iso{56}{}Ni$($\epsilon\gamma$)$\chem \iso{56}{}Co$($\epsilon\gamma,\beta^+\gamma$)$\chem \iso{56}{}Fe$, i.e., it does not occur at higher than initial abundances in stellar winds. This means that the yields of $\chem \iso{56}{}Fe$ (and other iron-group elements) from the pre-supernova evolution are negligible. 

The main isotopes of AGB nucleosynthesis (see also Table\,\ref{tab:yields1}) stem from the following reaction channels: the triple-alpha process that leads to carbon $\chem \iso{4}{}He$($\alpha,\gamma$)$\chem \iso{8}{}Be$($\alpha,\gamma$)$\chem \iso{12}{}C$; nitrogen production via proton capture $\chem \iso{17}{}O$($p,\alpha$)$\chem \iso{14}{}N$; oxygen by alpha capture, primarily 
through the channel $\chem \iso{12}{}C$($\alpha,\gamma$)$\chem \iso{16}{}O$. The created oxygen can be destroyed through $\chem \iso{16}{}O$($p,\gamma$)$\chem \iso{17}{}F$. Neon is produced via the reactions $\chem \iso{19}{}F$($\alpha,p$)$\chem \iso{22}{}Ne$ and $\chem \iso{14}{}N$($\alpha,\gamma$)$\chem \iso{18}{}F$($\beta^+,\nu$)$\chem \iso{18}{}O$($\alpha,\gamma$)$\chem \iso{22}{}Ne$, and sodium through proton capture on neon $\chem \iso{22}{}Ne$($p,\gamma$)$\chem \iso{23}{}Na$. The magnesium isotopes, created via alpha capture, $\chem \iso{22}{}Ne$($\alpha,n$)$\chem \iso{25}{}Mg$ and $\chem \iso{22}{}Ne$($\alpha,\gamma$)$\chem \iso{26}{}Mg$, have similar reaction rates at the stellar energies of AGB stars and the pre-supernova evolution of massive stars (e.g. \citealp{KAR10, DOH14}).

\subsubsection{Quantitative yields} \label{sect:enrichment_quantitative}

A SSE-synthesised $12$\,Gyr ($Z=0.004$) population will not contain any stars of $M_{\rm init}\ga0.9$\,M$_{\odot}$ while a $3$\,Gyr ($Z=0.008$) population will not have stars of $M_{\rm init}\ga1.4$\,M$_{\odot}$. Of a subset of the young population, $0-3$\,Myr, all stellar masses are expected to be present; of a subset of $60$\,Myr ($Z=0.02$) stars only, stars of $M_{\rm init}\ga6.4$\,M$_{\odot}$ will be absent (see also Section\,\ref{sect:distr_young}).\footnote{Similar results are obtained by the classical analytical approximation which gives a lifetime of a star (in yr) as $10^{10}/M_{\rm init}^{s-1}$, where $s=3$ for massive stars ($M_* > 30$\,M$_{\odot})$ and $s=4$ for lower-mass stars.} Note that main-sequence O and B stars, almost certainly contained in the $0-60$\,Myr group, despite being important energetically, do not significantly add to the nucleosynthetic yields. 

From a slightly different perspective, we can say that among the still living stars of $12$\,Gyr ($Z=0.004$), there will be stars in the current jet of the initial mass range $0.08\le M_{\rm init}\le0.9$\,M$_{\odot}$; this will contain a group of AGB stars of only $M_{\rm init}\sim0.9$\,M$_{\odot}$. Among the living stars of $3$\,Gyr ($Z=0.008$), there will be stars of the range $0.08\le M_{\rm init}\le1.4$\,M$_{\odot}$; amidst these there will be solely AGB stars of $M_{\rm init}\sim1.4$\,M$_{\odot}$. Among the living stars of $60$\,Myr ($Z=0.02$), there will be stars of range $0.08\le M_{\rm init}\le6$\,M$_{\odot}$; out of these, there will be AGB stars of only $M_{\rm init}\sim6$\,M$_{\odot}$.

Due to their high relative number, as argued in Section\,\ref{sect:eintercept}, and a longer mean stage duration (Table\,\ref{tab:para}), AGB stars are the main representative of the high-mass-loss stars in Centaurus\,A's jet plasma. The low fraction of young stars in the jet (alluded to in Sections\,\ref{sect:numberstars} and \ref{sect:spectrum_results}) causes the lower-mass AGB stars with $M_{\rm init}\sim0.9$\,M$_{\odot}$ and $M_{\rm init}\sim1.4$\,M$_{\odot}$ to numerically dominate the jet-contained AGB population. Given the high percentage of $12$\,Gyr old stars overall (see Section\,\ref{sect:distr_old} and Table\,\ref{tab:para2}), AGB stars of $M_{\rm init}\sim0.9$\,M$_{\odot}$ are expected to be the foremost representative.

To calculate the net yields of individual stars we integrate the mass lost from the model star over the star's lifetime according to
\begin{equation}
M_{\rm y}(k) = \int_{0}^{\tau} \left[\,X(k) - X_{\rm init} (k)\,\right] \dot{M}\,{\rm d}t\,,    \label{eq:yield_individual}
\end{equation}
where $M_{\rm y}(k)$ is the yield of species $k$ (in M$_{\odot}$), $\dot{M}$ is the current mass-loss rate, $X(k)$ and $X_{\rm init} (k)$ refer to the current and initial mass fraction of species $k$, and $\tau$ is the total lifetime of the stellar model. The net yield is positive, in the case where the element is produced (e.g. $\chem \iso{4}{}He$) and negative, if it is destroyed (e.g. $\chem \iso{16}{}O$). The $Z = 0.004$, $0.008$ and $0.02$ nucleosynthetic yields of $M_{\rm init}=1-6$\,M$_{\odot}$ stars have been computed by \cite{KAR10}, with the yields from a $M_{\rm init}=0.9$\,M$_{\odot}$, $Z = 0.004$ model calculated specifically for this paper. The $M_{\rm init}=1.4$\,M$_{\odot}$, $Z = 0.008$ yields are technically for a $M_{\rm init}=1.5$\,M$_{\odot}$, $Z = 0.008$ model since we do not have a $M_{\rm init}=1.4$\,M$_{\odot}$, $Z = 0.008$ model; however, the differences are expected to be very small and so we can use the yields as representative of a model of this mass.

The resultant net yields, $M_{\rm y}(k)$, for our adopted stellar populations are tabulated in Table\,\ref{tab:yields1}, i. e. yields with respect to the initial composition, whereas the absolute yields, i.e. the amount of material actually expelled, are given in Table\,\ref{tab:yields2}. Both include the contribution from the RGB through to the tip of the AGB phase. However, the yields are weighted towards the AGB as these are associated with the strongest stellar wind and the greatest mass loss. Mass lost during the preceding RGB stage is relatively small compared to the AGB phase, except for the lowest-mass stars that enter the AGB ($M_{\rm init}\sim0.9$\,M$_{\odot}$ stars). To show yields for a range of stellar ages (we would ideally compute a range $0-60$\,Myr for the young component), we would need a chemical evolution model, which is beyond the scope of this paper. One can see (Table\,\ref{tab:yields2}) that $\chem \iso{4}{}He$ is over-abundant relative to other nucleosynthetic products by a factor $\ga138$ ($12$\,Gyr $Z=0.004$ stars), a factor $\ga72$ ($3$\,Gyr $Z=0.008$ stars) and $\ga44$ ($60$\,Myr $Z=0.02$ stars). $\chem \iso{14}{}N$ is slightly under-abundant to $\chem \iso{12}{}C$, except for the case of young stars. $\chem \iso{16}{}O$ shows high absolute values since it is copiously present in the initial composition; it is not generally produced in low-mass stars during the AGB phase of evolution (\citealp{KAR10}, but see \citealp{PIG13} who illustrate that different treatment of convection can lead to $\chem \iso{16}{}O$ production). Intermediate-mass stars that experience hydrogen burning at the base of the envelope will destroy $\chem \iso{16}{}O$ (see Section\,\ref{sect:enrichment_general}). 
\begin{table}
\begin{center}
\caption{Net nucleosynthetic yields (in M$_{\odot}$) for the most abundant isotopes. The amount expelled per model star (by RGB to AGB evolutionary phases) into the jet by $12$\,Gyr ($Z=0.004$), $3$\,Gyr ($Z=0.008$) and $60$\,Myr ($Z=0.02$) stars.}
\label{tab:yields1}
\begin{tabular}{lccc}
\hline \\ [-1.5ex]
 Isotope  & & Amount expelled  \\
  & age $=12$\,Gyr & age $=3$\,Gyr & age $=60$\,Myr \\  
  & $Z=0.004$ & $Z=0.008$ & $Z=0.02$ \\
  & $0.9$\,M$_{\odot}$\,stars & $1.4$\,M$_{\odot}$\,stars & $6$\,M$_{\odot}$\,stars\\
[-1.5ex] 
\multicolumn{4}{l}{}\\
\hline \\ [-1.5ex]
$\chem \iso{1}{}H$   & $-4.72\times10^{-3}$ & $-1.74\times10^{-2}$ & $-3.52\times10^{-1}$ \\
$\chem \iso{3}{}He$  & $8.67\times10^{-5}$  & $2.94\times10^{-4}$  & $4.29\times10^{-6}$ \\
$\chem \iso{4}{}He$  & $4.63\times10^{-3}$  & $1.68\times10^{-2}$  & $3.31\times10^{-1}$ \\
$\chem \iso{12}{}C$  & $-2.25\times10^{-5}$ & $-1.04\times10^{-4}$ & $-9.86\times10^{-3}$ \\
$\chem \iso{13}{}C$  & $3.32\times10^{-6}$  & $1.96\times10^{-5}$  & $5.55\times10^{-4}$ \\
$\chem \iso{14}{}N$  & $2.54\times10^{-5}$  & $3.93\times10^{-4}$  & $3.62\times10^{-2}$ \\
$\chem \iso{16}{}O$  & $-2.69\times10^{-6}$ & $-8.42\times10^{-6}$ & $-8.03\times10^{-3}$ \\
$\chem \iso{20}{}Ne$ & $4.23\times10^{-8}$  & $-1.12\times10^{-7}$ & $-1.37\times10^{-5}$ \\
$\chem \iso{22}{}Ne$ & $-1.11\times10^{-7}$ & $5.76\times10^{-6}$  & $8.98\times10^{-4}$ \\
$\chem \iso{23}{}Na$ & $1.31\times10^{-7}$  & $1.54\times10^{-5}$  & $1.31\times10^{-4}$ \\
$\chem \iso{24}{}Mg$ & $1.73\times10^{-8}$  & $1.58\times10^{-8}$  & $-5.11\times10^{-5}$ \\
$\chem \iso{25}{}Mg$ & $-8.30\times10^{-9}$ & $-5.48\times10^{-8}$ & $1.27\times10^{-4}$ \\
$\chem \iso{26}{}Mg$ & $6.79\times10^{-9}$  & $1.07\times10^{-8}$  & $2.47\times10^{-4}$ \\
$\chem \iso{27}{}Al$ & $4.52\times10^{-9}$  & $6.45\times10^{-8}$  & $2.12\times10^{-5}$ \\
$\chem \iso{28}{}Si$ & $2.16\times10^{-8}$  & $1.66\times10^{-8}$  & $1.30\times10^{-5}$ \\
$\chem \iso{32}{}S$  & $1.31\times10^{-8}$  & $7.80\times10^{-9}$  & $-5.69\times10^{-6}$ \\
$\chem \iso{34}{}S$  & $6.18\times10^{-9}$  & $4.50\times10^{-9}$  & $5.61\times10^{-8}$ \\
$\chem \iso{56}{}Fe$ & $3.87\times10^{-8}$  & $1.0\times10^{-8}$   & $-4.10\times10^{-5}$ \\
[1ex] \hline
\end{tabular} \\
\begin{tabular}{l}
\end{tabular}
\end{center}
\end{table}
\begin{table}
\begin{center}
\caption{Absolute nucleosynthetic yields (in M$_{\odot}$) for the most abundant isotopes. The amount expelled per model star (by RGB to AGB evolutionary phases) into the jet by $12$\,Gyr ($Z=0.004$), $3$\,Gyr ($Z=0.008$) and $60$\,Myr ($Z=0.02$) stars.}
\label{tab:yields2}
\begin{tabular}{lccc}
\hline \\ [-1.5ex]
 Isotope  & & Amount expelled  \\
  & age $=12$\,Gyr & age $=3$\,Gyr & age $=60$\,Myr \\  
  & $Z=0.004$ & $Z=0.008$ & $Z=0.02$ \\
  & $0.9$\,M$_{\odot}$\,stars & $1.4$\,M$_{\odot}$\,stars & $6$\,M$_{\odot}$\,stars\\
[-1.5ex] 
\multicolumn{4}{l}{}\\
\hline \\ [-1.5ex]
$\chem \iso{1}{}H$   & $2.18\times10^{-1}$ & $6.22\times10^{-1}$ & $3.14$ \\
$\chem \iso{3}{}He$  & $8.67\times10^{-5}$ & $2.94\times10^{-4}$ & $4.29\times10^{-6}$ \\
$\chem \iso{4}{}He$  & $7.82\times10^{-2}$ & $2.38\times10^{-1}$ & $1.82$ \\
$\chem \iso{12}{}C$  & $1.58\times10^{-4}$ & $9.47\times10^{-4}$ & $5.53\times10^{-3}$ \\
$\chem \iso{13}{}C$  & $5.49\times10^{-6}$ & $3.23\times10^{-5}$ & $7.40\times10^{-4}$ \\
$\chem \iso{14}{}N$  & $9.11\times10^{-5}$ & $7.76\times10^{-4}$ & $4.18\times10^{-2}$ \\
$\chem \iso{16}{}O$  & $5.67\times10^{-4}$ & $3.32\times10^{-3}$ & $4.07\times10^{-2}$ \\
$\chem \iso{20}{}Ne$ & $9.63\times10^{-5}$ & $5.62\times10^{-4}$ & $8.21\times10^{-3}$ \\
$\chem \iso{22}{}Ne$ & $7.63\times10^{-6}$ & $5.10\times10^{-5}$ & $1.56\times10^{-3}$ \\
$\chem \iso{23}{}Na$ & $2.12\times10^{-6}$ & $1.31\times10^{-5}$ & $3.01\times10^{-4}$ \\
$\chem \iso{24}{}Mg$ & $3.06\times10^{-5}$ & $1.79\times10^{-4}$ & $2.56\times10^{-3}$ \\
$\chem \iso{25}{}Mg$ & $4.01\times10^{-6}$ & $2.34\times10^{-5}$ & $4.71\times10^{-4}$ \\
$\chem \iso{26}{}Mg$ & $4.62\times10^{-6}$ & $2.70\times10^{-5}$ & $6.41\times10^{-4}$ \\
$\chem \iso{27}{}Al$ & $3.46\times10^{-6}$ & $2.02\times10^{-5}$ & $3.16\times10^{-4}$ \\
$\chem \iso{28}{}Si$ & $3.89\times10^{-5}$ & $2.27\times10^{-4}$ & $3.33\times10^{-3}$ \\
$\chem \iso{32}{}S$  & $2.36\times10^{-5}$ & $1.37\times10^{-4}$ & $2.01\times10^{-3}$ \\
$\chem \iso{34}{}S$  & $1.11\times10^{-5}$ & $6.49\times10^{-5}$ & $9.49\times10^{-4}$ \\
$\chem \iso{56}{}Fe$ & $6.96\times10^{-5}$ & $4.06\times10^{-4}$ & $5.90\times10^{-3}$ \\
[1ex] \hline
\end{tabular} \\
\begin{tabular}{l}
\end{tabular}
\end{center}
\end{table}

We use $\chem \iso{12}{}C$ as an example for the calculation of the amount of a specific isotope lost
\begin{table}
\begin{center}
\caption{Mass-loss rates in individual isotopes (in M$_{\odot}$\,yr$^{-1}$) for the most abundant nucleosynthetic species. The amount per year released to the jet by $75$ per cent of $12$\,Gyr ($Z=0.004$), $25$ per cent of $3$\,Gyr ($Z=0.008$) and $0-0.5$ per cent of $60$\,Myr ($Z=0.02$) AGB stars. AGB phase durations and the percentage of material lost are, respectively, $\sim21.8$\,Myr and $\sim42.4$ per cent (for $M_{\rm init}=0.9$\,M$_{\odot}$, $Z=0.004$ stars), $\sim15.5$\,Myr and $\sim87.8$ per cent (for $M_{\rm init}=1.4$\,M$_{\odot}$, $Z=0.008$ stars) and $\sim1.21$\,Myr and $\sim99.3$ per cent (for $M_{\rm init}=6$\,M$_{\odot}$, $Z=0.02$ stars). See the main text for an example calculation.}
\label{tab:yields3}
\begin{tabular}{lccc}
\hline \\ [-1.5ex]
 Isotope  & & Amount expelled  \\
  & age $=12$\,Gyr & age $=3$\,Gyr & age $=60$\,Myr \\  
  & $Z=0.004$ & $Z=0.008$ & $Z=0.02$ \\
  & $0.9$\,M$_{\odot}$\,stars & $1.4$\,M$_{\odot}$\,stars & $6$\,M$_{\odot}$\,stars\\
  & $75\%$                   & $25\%$                    & $0-0.5\%$\\
[-1.5ex] 
\multicolumn{4}{l}{}\\
\hline \\ [-1.5ex] 
$\chem \iso{1}{}H$   & $3.17\times10^{-5}$  & $8.80\times10^{-5}$ & $0-1.29\times10^{-4}$ \\
$\chem \iso{3}{}He$  & $1.27\times10^{-8}$  & $4.16\times10^{-8}$ & $0-1.76\times10^{-10}$ \\
$\chem \iso{4}{}He$  & $1.14\times10^{-5}$  & $3.37\times10^{-5}$ & $0-7.46\times10^{-5}$ \\
$\chem \iso{12}{}C$  & $2.30\times10^{-8}$  & $1.34\times10^{-7}$ & $0-2.27\times10^{-7}$ \\
$\chem \iso{14}{}N$  & $1.33\times10^{-8}$  & $1.10\times10^{-7}$ & $0-1.72\times10^{-6}$ \\
$\chem \iso{16}{}O$  & $8.28\times10^{-8}$  & $4.70\times10^{-7}$ & $0-1.67\times10^{-6}$ \\
$\chem \iso{20}{}Ne$ & $1.40\times10^{-8}$  & $7.96\times10^{-8}$ & $0-3.37\times10^{-7}$ \\
$\chem \iso{22}{}Ne$ & $1.11\times10^{-9}$  & $7.22\times10^{-9}$ & $0-6.40\times10^{-8}$ \\
$\chem \iso{24}{}Mg$ & $4.47\times10^{-9}$  & $2.53\times10^{-8}$ & $0-1.05\times10^{-7}$ \\
$\chem \iso{26}{}Mg$ & $6.74\times10^{-10}$ & $3.82\times10^{-9}$ & $0-2.63\times10^{-8}$ \\
$\chem \iso{28}{}Si$ & $5.67\times10^{-9}$  & $3.21\times10^{-8}$ & $0-1.37\times10^{-7}$ \\
$\chem \iso{32}{}S$  & $8.10\times10^{-9}$  & $1.95\times10^{-8}$ & $0-8.23\times10^{-8}$ \\
$\chem \iso{56}{}Fe$ & $1.02\times10^{-8}$  & $5.75\times10^{-8}$ & $0-2.42\times10^{-7}$ \\
[1ex] \hline
\end{tabular} \\
\begin{tabular}{l}
\end{tabular}
\end{center}
\end{table}
per year, for the most recent year of the $12$\,Gyr ($Z = 0.004$) stars. The total amount lost during the $M_{\rm init}=0.9$\,M$_{\odot}$ model's lifetime is $\sim 1.6\times10^{-4}$\,M$_{\odot}$ (Table\,\ref{tab:yields2}). In total, $\sim2.8\times10^{-1}$\,M$_{\odot}$ is released, with $\sim42$ per cent lost during the AGB. Hence, roughly, $6.7\times10^{-5}$\,M$_{\odot}$ of $\chem \iso{12}{}C$ is lost during the AGB. Given the duration of the AGB phase of a $M_{\rm init}\sim0.9$\,M$_{\odot}$ ($Z = 0.004$) star of $\sim21.8$\,Myr, this means a mass-loss rate in $\chem \iso{12}{}C$ of $\dot{M}_{\chem \iso{12}{}C} \sim3.1\times10^{-12}$\,M$_{\odot}$\,yr$^{-1}$. We next multiply this by the number of specific AGB stars in the jet. Of the $\sim1\times10^4$ AGB stars estimated in the jet (Section\,\ref{sect:eintercept}), $75$ per cent of them, i.e. $7.5\times10^3$, are $M_{\rm init}\sim0.9$\,M$_{\odot}$ ($Z = 0.004$) stars, $25$ per cent, i.e. $2.5\times10^3$, are $M_{\rm init}\sim1.4$\,M$_{\odot}$ ($Z = 0.008$) stars, and $0-0.5$ per cent, i.e. $0-50$, are $M_{\rm init}\sim6$\,M$_{\odot}$ ($Z = 0.02$). This leads to $\dot{M}_{\chem \iso{12}{}C} \sim2.3\times10^{-8}$\,M$_{\odot}$\,yr$^{-1}$ for the ensemble of $12$\,Gyr ($Z = 0.004$) stars (see Table\,\ref{tab:yields3}). Over the probable physical lifetime of the current jet ($\sim2$\,Myr) and the giant lobes ($\sim560$\,Myr), this is a $\chem \iso{12}{}C$ mass of, respectively, $\sim4.6\times10^{-2}$ and $\sim13$\,M$_{\odot}$ (i.e. $0.001$ per cent of the all-particle mass lost). The results for other, main isotopes and other AGB stars (AGB phase durations of $\sim15.5$\,Myr for the $M_{\rm init}=1.4$\,M$_{\odot}$ $Z = 0.008$ component and $\sim1.2$\,Myr for $M_{\rm init}=6$\,M$_{\odot}$ $Z = 0.02$) are summarized in Table\,\ref{tab:yields3}.

The remainder, i.e. $8\times10^8$ (total) $- 1\times10^4$ (AGB) stars, are assumed to be on the main sequence. $75$ per cent of this, i.e. $6\times10^8$ stars, are of the $12$\,Gyr ($Z = 0.004$) population, $25$ per cent, i.e. $2\times10^8$ stars, are of $3$\,Gyr ($Z = 0.008$) and $0-0.5$ per cent, i.e. $0-4\times10^6$ stars, are of $60$\,Myr ($Z = 0.02$). We adopt a solar-like main-sequence mass loss of $\sim2\times10^{-14}$\,M$_{\odot}$\,yr$^{-1}$ for each of these stars, which gives $\dot{M}\sim1.6\times10^{-5}$\,M$_{\odot}$\,yr$^{-1}$ overall for the all-particle main-sequence loss. To calculate the composition of that material, we need to break down this mass-loss rate into $75$ per cent of $12$\,Gyr ($Z=0.004$), $25$ per cent of $3$\,Gyr ($Z=0.008$) and a few tens of per cent of $60$\,Myr ($Z=0.02$). Assuming a scaled-solar isotopic breakdown: $\chem \iso{4}{}He$ is initially $\sim0.25$ in a $Z =0.004$ population; this yields $\dot{M}_{\chem \iso{4}{}He}\sim0.25\times1.2\times10^{-5}\sim3.0\times10^{-6}$\,M$_{\odot}$\,yr$^{-1}$. For other isotopes, e.g., $\chem \iso{12}{}C$, the calculation is similar: the mass fraction is initially $X(\chem \iso{12}{}C)\sim2.5\times10^{-4}\times12\times1.2\times10^{-5}$, divided by ($0.02/0.004$) to get an initial scaled composition at $Z = 0.004$, from which follows $\dot{M}_{\chem \iso{12}{}C}\sim7.2\times10^{-9}$\,M$_{\odot}$\,yr$^{-1}$. 
\begin{table}
\begin{center}
\caption{Mass-loss rates in individual isotopes (in M$_{\odot}$\,yr$^{-1}$) for the most abundant nucleosynthetic species. The rough amount per year released to the jet by $75$ per cent of $12$\,Gyr ($Z=0.004$), $25$ per cent of $3$\,Gyr ($Z=0.008$) and $0-0.5$ per cent of $60$\,Myr ($Z=0.02$) main-sequence stars. See the main text for an example calculation.}
\label{tab:yields4}
\begin{tabular}{lccc}
\hline \\ [-1.5ex]
 Isotope  & & Amount expelled  \\
  & age $=12$\,Gyr & age $=3$\,Gyr & age $=60$\,Myr \\  
  & $Z=0.004$ & $Z=0.008$ & $Z=0.02$ \\
  & $0.08-0.85$\,M$_{\odot}$ & $0.08-1.35$\,M$_{\odot}$ & $0.08-5.55$\,M$_{\odot}$\\
  & $75\%$                   & $25\%$                    & $0-0.5\%$\\
[-1.5ex] 
\multicolumn{4}{l}{}\\
\hline \\ [-1.5ex]
$\chem \iso{1}{}H$   & $8.9\times10^{-6}$  & $ 2.9\times10^{-6}$ & $0-5.6\times10^{-8}$ \\
$\chem \iso{3}{}He$  & $7.0\times10^{-10}$ & $4.7\times10^{-10}$ & $0-2.3\times10^{-11}$ \\
$\chem \iso{4}{}He$  & $2.9\times10^{-6}$  & $1.0\times10^{-6}$  & $0-2.2\times10^{-8}$ \\
$\chem \iso{12}{}C$  & $7.2\times10^{-9}$  & $4.8\times10^{-9}$  & $0-2.4\times10^{-10}$ \\
$\chem \iso{14}{}N$  & $2.6\times10^{-9}$  & $1.8\times10^{-9}$  & $0-8.8\times10^{-11}$ \\
$\chem \iso{16}{}O$  & $2.3\times10^{-8}$  & $1.5\times10^{-8}$  & $0-7.6\times10^{-10}$ \\
$\chem \iso{20}{}Ne$ & $2.9\times10^{-9}$  & $2.6\times10^{-9}$  & $0-1.3\times10^{-10}$ \\
$\chem \iso{22}{}Ne$ & $3.1\times10^{-10}$ & $2.1\times10^{-10}$ & $0-1.0\times10^{-11}$ \\
$\chem \iso{24}{}Mg$ & $1.2\times10^{-9}$  & $8.2\times10^{-10}$ & $0-4.1\times10^{-11}$ \\
$\chem \iso{26}{}Mg$ & $1.9\times10^{-10}$ & $1.2\times10^{-10}$ & $0-6.2\times10^{-12}$ \\
$\chem \iso{28}{}Si$ & $1.6\times10^{-9}$  & $1.0\times10^{-9}$  & $0-5.2\times10^{-11}$ \\
$\chem \iso{32}{}S$  & $9.5\times10^{-10}$ & $6.3\times10^{-10}$ & $0-3.2\times10^{-11}$ \\
$\chem \iso{56}{}Fe$ & $2.8\times10^{-9}$  & $1.9\times10^{-9}$  & $0-9.3\times10^{-11}$ \\
[1ex] \hline
\end{tabular} \\
\begin{tabular}{l}
\end{tabular}
\end{center}
\end{table}
\begin{table}
\begin{center}
\caption{Approximate amount of mass (in M$_{\odot}$) in individual isotopes lost by the combined AGB and main-sequence phases to the current- and the pre-existing jet (expressed by the physical age of the giant lobes) by the ensemble of $75$ per cent of $12$\,Gyr ($Z=0.004$), $25$ per cent of $3$\,Gyr ($Z=0.008$) and $0-0.5$ per cent of $60$\,Myr ($Z=0.02$) stars.}
\label{tab:yields5}
\begin{tabular}{lccc}
\hline \\ [-1.5ex]
 Isotope  &\hspace{1.9cm} Amount expelled \\
  & current jet & pre-existing jet \\ 
  & $2$\,Myr & $560$\,Myr \\  
[-1.5ex] 
\multicolumn{3}{l}{}\\
\hline \\ [-1.5ex]
$\chem \iso{1}{}H$   & $2.6\times10^2$    & $7.4\times10^4$ \\
$\chem \iso{3}{}He$  & $1.1\times10^{-1}$ & $3.1\times10^1$ \\
$\chem \iso{4}{}He$  & $9.8\times10^1$    & $2.7\times10^4$ \\
$\chem \iso{12}{}C$  & $3.4\times10^{-1}$ & $9.5\times10^1$ \\
$\chem \iso{14}{}N$  & $2.6\times10^{-1}$ & $7.2\times10^1$ \\
$\chem \iso{16}{}O$  & $1.2$              & $3.3\times10^2$ \\
$\chem \iso{20}{}Ne$ & $2.0\times10^{-1}$ & $5.5\times10^1$ \\
$\chem \iso{22}{}Ne$ & $1.8\times10^{-2}$ & $5.0$ \\
$\chem \iso{24}{}Mg$ & $6.4\times10^{-2}$ & $1.8\times10^1$ \\
$\chem \iso{26}{}Mg$ & $9.6\times10^{-3}$ & $2.7$ \\
$\chem \iso{28}{}Si$ & $8.1\times10^{-2}$ & $2.3\times10^1$ \\
$\chem \iso{32}{}S$  & $5.8\times10^{-2}$ & $1.6\times10^1$ \\
$\chem \iso{56}{}Fe$ & $1.5\times10^{-1}$ & $4.1\times10^1$ \\
[1ex] \hline
\end{tabular} \\
\begin{tabular}{l}
\end{tabular}
\end{center}
\end{table}
The main-sequence mass-loss rates in individual isotopes are tabulated in Table\,\ref{tab:yields4}. Finally, Table\,\ref{tab:yields5} shows the approximate amount of mass in individual isotopes lost by the combined AGB and main-sequence phases to the current jet, and also to the older jet that inflated the giant lobes (which is of a similar or slightly higher jet power, $1-5\times10^{43}$\,erg\,s$^{-1}$; \citealp{WYK13}). This indicates that the most abundant nuclei make up $\sim2.1$ per cent ($\chem \iso{4}{}He$), $\sim0.026$ per cent ($\chem \iso{16}{}O$), $\sim0.007$ per cent ($\chem \iso{12}{}C$), $\sim0.006$ per cent ($\chem \iso{14}{}N$) and $\sim0.004$ per cent ($\chem \iso{20}{}Ne$) of the total mass in the jet and the lobes.

To gauge the composition of the jet (and lobes, to which these isotopes will be passed) in terms of the entire potential stellar content, we would need to add the ejecta from stars in the range $8-10$\,M$_{\odot}$ (super-AGB and low-mass stars that become supernovae) as well as more massive stars that become WRs and higher-mass supernovae. Recall that pre-supernova phases of stellar evolution are predicted to expel a large amount of $\chem \iso{16}{}O$ (e.g. \citealp{HIR05, CHI13}; Section\,\ref{sect:enrichment_general}). Moreover, carbon-rich WR stars are thought to expel high quantities of $\chem \iso{22}{}Ne$ (e.g. \citealp{MAE93}). We do not model the pre-supernova evolution of massive stars here but the yields we calculate do at least provide an indication of the main-sequence phase and of the AGB contribution, which is likely to be of greatest importance overall (see above).

If some of the events at $\ge55$\,EeV registered by the Pierre Auger Observatory originate from internal entrainment in Centaurus\,A (in the jet with a subsequent transport to the giant lobes for the final, stochastic acceleration, as suggested by \citealp{WYK13}), the predominant UHECR composition at the detector from this source\footnote{$\chem \iso{4}{}He$ has relatively high yields but is, due to its low charge, probably not accelerated to the UHE regime in Centaurus\,A (see \citealp{WYK13}). Furthermore, due to quasi-deuteron excitation, $\chem \iso{4}{}He$ is expected to photodisintegrate on the distance to this source.} is expected to be a mixture of $\chem \iso{16}{}O$\,/\,$\chem \iso{12}{}C$\,/\,$\chem \iso{14}{}N$\,/\,$\chem \iso{20}{}Ne$\,/\,$\chem \iso{56}{}Fe$\,/\,$\chem \iso{28}{}Si$\,/\,$\chem \iso{24}{}Mg$\,/\,$\chem \iso{26}{}Mg$, with $\chem \iso{16}{}O$, $\chem \iso{12}{}C$ and $\chem \iso{14}{}N$ the key isotopes. Thus the scattering for these UHECRs in the intergalactic and Galactic magnetic field may well be substantial. For more distant FR\,I sources, the UHECRs accelerated at the source will photodisintegrate {\it en route} to an Earth-based detector,\footnote{Photodisintegration is a gradual process; it takes several steps to break down the original nucleus to protons.} showing a lightening in the composition compared to the one at the source of origin (e.g. \citealp{ALL05}).

\section{Summary and Conclusions} \label{sect:summary}

We have modelled mass loading through stellar winds of Centaurus\,A's jet using stellar evolution- and wind codes by \cite{HUR00}, \cite{CRA11} and \cite{VIN99, VIN00, VIN01}, and computed conjointly the entrainment rates and nucleosynthetic isotope yields. The principal novelties of this paper are a better estimate of the mass input rate by using more realistic stellar populations, an estimate of particle acceleration luminosity and spectrum, and abundances of the entrained material. The key results are as follows:

(1) From $R$-band photometry and a SSE-synthesised NGC\,5128's stellar population with ages and metallicities as $12$\,Gyr at $Z=0.004$, $3$\,Gyr at $Z=0.008$ and $0-60$\,Myr at $Z=0.02$, we infer $\sim8\times10^8$ for the total number of stars within the jet volume. An obvious observational proxy for the fraction of the $0-60$\,Myr stars in the jet is lacking; based on our modelling, we advance the possibility that this fraction is $0-0.5$ per cent.

(2) Energetically, we can meet the $E_{\rm intercept,X}$ criterion: our jet-stellar wind interaction model, which relies on Fermi\,I-type particle acceleration, produces X-rays, even for zero fraction young stars. The model can reproduce the combined diffuse- and knot X-ray luminosity of the whole $4.5$\,kpc-scale jet of Centaurus\,A of $\sim1\times10^{39}$\,erg\,s$^{-1}$. We also produce the broad-band spectrum of the kpc jet up to the optical, albeit in a region outside the starburst that might be expected to have fewer young stars. We recover the mean X-ray spectral index for sensible fractions of young stars of order $0-0.5$ per cent. Given the stellar age constraints and the plausible fraction of young stars, the AGB stars must numerically dominate over their high-mass-loss counterparts currently present in the jet; among the AGB stars, those with $M_{\rm init}\sim0.9$\,M$_{\odot}$ ought to be the foremost representative. 

(3) We propose that the jet experiences increasing baryon fraction and derive an entrainment rate of $\sim2.3\times10^{-3}$\,M$_{\odot}$\,yr$^{-1}$, which is within a factor $\sim2$ of the rough estimate of internal entrainment rate by \cite{WYK13}. Such an amount of material can cause substantial deceleration, by virtue of momentum balance, of the present-day jet.

(4) We have established that AGB stars of $12$\,Gyr ($Z=0.004$), $3$\,Gyr ($Z=0.008$) and $60$\,Myr ($Z=0.02$) principally contribute towards $\chem \iso{4}{}He$, $\chem \iso{16}{}O$, $\chem \iso{12}{}C$, $\chem \iso{14}{}N$ and $\chem \iso{20}{}Ne$ nuclei in the jet. As `super-AGB' stars mainly produce $\chem \iso{4}{}He$, $\chem \iso{14}{}N$, $\chem \iso{25}{}Mg$ and $\chem \iso{26}{}Mg$, and main-sequence and pre-supernova phases add a large fraction of $\chem \iso{16}{}O$, we predict that, if some of the Auger Observatory events of $\ge55$\,EeV originate from internal entrainment in Centaurus\,A, their composition is plausibly predominantly $\chem \iso{16}{}O$\,/\,$\chem \iso{12}{}C$\,/\,$\chem \iso{14}{}N$\,/\,$\chem \iso{20}{}Ne$\,/\,$\chem \iso{56}{}Fe$\,/\,$\chem \iso{28}{}Si$\,/\,$\chem \iso{24}{}Mg$\,/\,$\chem \iso{26}{}Mg$, with $\chem \iso{16}{}O$, $\chem \iso{12}{}C$ and $\chem \iso{14}{}N$ being the key elements.\footnote{Recently posted results from the Pierre Auger Observatory \citep{AAB14} provide a strong indication for a mixed particle composition at the detector, with a prominent role for intermediate-mass nuclei. Whether this reflects the original source composition from a nearby source/nearby sources, photodisintegration products from a more distant one/ones, or both, remains to be answered.}

Targeting the unobscured knots in the Centaurus\,A's jet with the X-Shooter instrument on the VLT or with the future E-ELT, to search for $\chem \iso{16}{}O$ and $\chem \iso{12}{}C$ emission line spectra characteristic of strong stellar winds, may be a real test of the existence of young high-mass-loss stars in the jet. Further refinement of the star number and star population ages in Centaurus\,A's jet and the associated metallicities would put tighter constraints on the broad-band synchrotron spectrum of the jet. In a future paper, we will report on VLBA circular polarization observations designed to constrain the particle composition -- `light' (electron-positron) or `heavy' (electron-hadron) jet -- on the smallest scales.

\section*{Acknowledgements} 
We thank Marina Rejkuba, Frank Israel, Robert Laing, Diederik Kruijssen, Tom Jones, Onno Pols, Marc Sarzi, Peter Biermann, Lex Kaper, Olga Hartoog, Carola Dobrigkeit Chinellato, Huib Henrichs and Karl-Heinz Kampert for valuable discussions, and an anonymous referee for a thoughtful report. SW is grateful for hospitality at the University of Hertfordshire. This work has made use of the University of Hertfordshire Science and Technology Research Institute high-performance computing facility. AIK is supported through an Australian Research Council Future Fellowship (FT110100475).

\bsp

\label{lastpage}

\end{document}